\documentstyle[epsfig]{mn}
\begin{document}
\input BoxedEPS.tex
\newcommand{\aip}{{\small ${\cal AIPS}$}}
\newcommand{\gtsim}{\mbox{{\raisebox{-0.4ex}{$\stackrel{>}{{\scriptstyle\sim}}
$}}}}
\newcommand{\ltsim}{\mbox{{\raisebox{-0.4ex}{$\stackrel{<}{{\scriptstyle\sim}}
$}}}}
\newcommand{\s}{$\stackrel{\rm s}{.}$}
\newcommand{\h}{$^{\rm h}$}
\newcommand{\m}{$^{\rm m}$}
\newcommand{\pp}{$\stackrel{\prime\prime}{.}$}
\newcommand{\de}{$^{\circ}$}
\newcommand{\p}{$^{\prime}$}
\newcommand{\arc}{$^{\prime\prime}$}
\newcommand{\marc}{^{\prime\prime}}
\newcommand{\rs}{{\em $r_s$}}
\newcommand{\DPM}{{\em DPM}}
\newcommand{\alf}{{\displaystyle\biggl({\nu_{\rm h} \over \nu_{\rm l}}\biggr)^{\alpha}} }

\newcommand{\figstart}[1]
    { \begin{figure}[htb]
      \begin{picture}(0,#1) }
\newcommand{\figend}[4]
    { \end{picture}
      \special{#1}
      \caption[#2]{#3}
      \label{#4}
      \end{figure} }
\newcommand{\fig}[5]
    { \figstart{#1}
      \figend{#2}{#3}{#4}{#5} }
\newcommand{\bHS}{\beta_{\mbox{\scriptsize HS}}}
\newcommand{\bBF}{\beta_{\mbox{\scriptsize BF}}}
\newcommand{\nT}{\nu_{\mbox{\scriptsize T}}}
\newcommand{\et}{E_{\mbox{\scriptsize T}}}
\newcommand{\nTn}{\nu_{\mbox{\scriptsize Tn}}}
\newcommand{\nTf}{\nu_{\mbox{\scriptsize Tf}}}
\newcommand{\tn}{\tau_{x\mbox{\scriptsize n}}}
\newcommand{\tf}{\tau_{x\mbox{\scriptsize f}}}
\newcommand{\xn}{x_{\mbox{\scriptsize n}}}
\newcommand{\xf}{x_{\mbox{\scriptsize f}}}
\newcommand{\yn}{y_{\mbox{\scriptsize n}}}
\newcommand{\yf}{y_{\mbox{\scriptsize f}}}
\newcommand{\lln}{l_{\mbox{\scriptsize n}}}
\newcommand{\llf}{l_{\mbox{\scriptsize f}}}
\newcommand{\Dn}{f(\Delta_{\mbox{\scriptsize n}})}
\newcommand{\Df}{f(\Delta_{\mbox{\scriptsize f}})}
\newcommand{\B}{\mbox{$B$}}
\newcommand{\Bo}{\mbox{$B$}_{0}}

\SetEPSFDirectory{/scratch/sbgs/figures/hst/}
\SetRokickiEPSFSpecial
\HideDisplacementBoxes

\title[Star formation history]{Quantifying dust and the ultraviolet radiation-density in the
local universe.}

\author[Rowan-Robinson M.]{Michael Rowan-Robinson\\
Astrophysics Group, Blackett Laboratory, Imperial College of Science Technology and Medicine,
\\
Prince Consort Road, London SW7 2BW}
\maketitle
\begin{abstract}
A sample of local galaxies for which far infrared and uv fluxes are available is used to
estimate the characteristic dust extinction in galaxies and to test whether standard
dust properties are plausible.  
Assuming galaxies can be characterized by a single dust optical depth (certainly not valid for galaxies
with a dominant starburst component), the infrared excess and ultraviolet
colours of local galaxies are found to be consistent with normal Milky Way dust, with a mean value for E(B-V)
of 0.16.  A significant fraction of the dust heating is due to older, lower mass stars, and
this fraction increases towards earlier galaxy types.

Analysis of $(F_{fir}/F_{uv})$ versus uv colour diagrams for starburst galaxies in terms of a simple
screen dust model does not support a Calzetti (1999) rather than a Milky Way extinction law, though the absence
of the expected strong 2200 $\AA$ feature in several galaxies with IUE spectra does show that  
more detailed radiative transfer models are needed, probably with non-spherical geometry.  

A simple treatment in which the 100/60 $\mu m$ flux-ratio is used to subtract the optically thick starburst
contribution to the far infrared radiation results in lower extinction estimates for the 
optically thin cirrus component, with a mean E(B-V) of 0.10

The uv luminosity density, corrected for dust extinction, is derived and a value for the local
mean star-formation rate inferred.  This is consistent with previous estimates from
uv surveys and from $H{\alpha}$ surveys.

\end{abstract}
\begin{keywords}
infrared: galaxies - galaxies: evolution - star:formation - galaxies: starburst - 
cosmology: observations
\end{keywords}


\section{Introduction}

The star formation history of the universe has attracted immense interest in recent years and has been 
studied out to redshifts of 5, using a variety of estimates of star
formation rate: ultraviolet, H$\alpha$, far infrared, submillimetre, radio 
(Lilley et al 1996, Madau et al 1996, Connelly et al 1997, Rowan-Robinson
et al 1997, Hughes et al 1998, Flores et al 1999, Steidel et al 1998, Cram 1998, Yan et al 1999, Haarsma et al 2000, 
Sullivan et al 2000, 2001, Hopkins et al 2001, Mann et al 2002).  However because of the strong
effects of absorption by insterstellar dust, the local star formation rate is not that well known.
It has been studied using H$\alpha$ by Gallego et al (1995), Gronwall (1998), Tresse and Maddox (1998),
and Sullivan et al (2001), and from uv data by Treyer et al (1998) and Sullivan et al (2000).  However the estimates
of dust extinction used in these studies are generally rather crude.  It has been argued that the best estimates of 
the local star-formation rate are therefore those derived from far infrared (IRAS) data, but  
even there we do not really know with precision what fraction of the far infrared radiation
is from dust illuminated by light from young, as opposed to old or middle-aged, stars, nor do we know
precisely what fraction of the light from young stars is absorbed by dust and what fraction escapes relatively
unscathed.

A further complication is doubt about the nature of the dust in regions of star-formation.  The absorption,
scattering and extinction properties of interstellar dust in our Galaxy are reasonably well-established
(Savage and Mathis 1979, Mathis 1990, see the recent comprehensive review by Calzetti 2001).
This standard interstellar extinction curve has been modelled successfully with multigrain
models (eg Draine and Lee 1984, Rowan-Robinson 1992, Siebenmorgen and Krugel 1992).  
Deviations from this standard interstellar extinction curve are found along different lines of sight in our Galaxy 
and in regions of low metallicity like the Magellanic Clouds (Nandy et al 1981).  However in a series of
 papers by Calzetti and her collaborators (Calzetti et al 1994, 1995, Calzetti 1997, Calzetti et al 1997, 
1999, 2000, Meurer et al 1999) the proposal
is made that the effective extinction law for starburst galaxies is different from that for the Milky Way.
Although Calzetti (2001) stresses that this is an empirical result, perhaps arising from dust geometry,
scattering etc, it is perplexing given the great success of models based on standard Milky Way dust in a
wide range of applications (eg Efstathiou and Rowan-Robinson 1995, Granato et al 1996, Silva et al 1998,
Efstathiou et al 2000, Granato et al 2001). 

In this paper I use a sample of very nearby galaxies ($V_o < 5000 km/s$) for which there are reliable far
infrared and far ultraviolet data to investigate the dust properties and magnitude of extinction in
normal and star-forming galaxies.  I investigate whether abnormal dust properties are required to understand
these galaxies and I estimate the local uv radiation-density, and hence the global star-formation rate at 
z = 0.  This same sample of galaxies has been used to study extinction in nearby galaxies by Xu and Buat (1995)
and Buat and Xu (1996).  Dust extinction in nearby galaxies has also been explored, using different
galaxy samples, by Meurer et al (1995, 1999), Xu et al (1997), Gordon et al (2000), Buat et al (2002), Goldader et al (2002), 
Bell et al (2002, 2003).  Most of these studies use the ratio of bolometric far infrared luminosity, $L_{FIR}$, to far uv 
luminosity, $L_{FUV}$ (usually at 1600 or 2000 $\AA$), to estimate extinction.  
The need for far ir data to help resolve ambiguities generated by the unknown dust
geometry has been emphasized by Witt and Gordon (2000), who find that the latter
has a strong impact on uv colours once $A_V >$ 0.5.
However as emphasized by Bell (2003)
use of $L_{FIR}/L_{FUV}$
on its own will not give an accurate estimate of the extinction because of the different relative
contributions of older stars to  $L_{FIR}$ in galaxies of different Hubble types.  Meurer et al (1999) find a strong
correlation of  $L_{FIR}/L_{FUV}$ with far ultraviolet continuum slope, $\beta$, for starburst galaxies, but 
Goldader et al (2002), Bell (2002) shows this does not hold for normal galaxies.  The contribution of older stars 
can be taken into account if an additional flux-ratio, for example
$L_B/L_{FUV}$ is brought into play (Buat and Xu 1996, Gordon et al 2000) and this is the method used here.  This does give a
self-consistent estimate of mean dust extinction.

However an additional complication not treated in previous studies is that the far infrared emission from normal galaxies 
is a mixture of emission from relatively optically thin dust associated with HI clouds, absorbing the general stellar
radiation field (the infrared cirrus), and emission from highly optically thick dust associated with molecular hydrogen
clouds illuminated by newly formed stars (starbursts) (Rowan-Robinson and Crawford 1989).  In this paper I use
the 100/60 $\mu m$ flux-ratio to separate these components.

The layout of the paper is as follows.  In section 2 I define a sample of
local galaxies for which both far infrared data is available from IRAS,
and large-beam ultraviolet data is available form balloon-borne studies,
to estimate the characteristic dust extinction in the ultraviolet and
correlate it with the blue-uv colour.  When allowance is made for
galaxy type, standard dust is consistent with the optical depth-
colour relations, even for starburst galaxies, in contrast to earlier claims that new dust properties are implied.

In section 3 I include the effect of separating the far infrared emission into cirrus and
starburst components.  In section 4 I derive the local, extinction-corrected, uv luminosity function and
give a new uv estimate of the local star formation rate.  Section 5 gives discussion and conclusions.

A Hubble constant of 100 $km/s/Mpc$ is used throughout.

\section{Dust extinction estimates for a local galaxy sample}

To try to shed light on this question of how much of the ultraviolet and visible 
light generated locally in star-forming regions is absorbed by dust, I have assembled a 
sample of 147 bright (B $\leq$ 13.8) nearby (V $<$ 5000 km$/$s), optically selected galaxies 
for which there are both 60 $\mu$m (S(60) $\geq$ 0.6 Jy) and large-beam 2000 $\AA$ observations
(Donas et al 1987, Deharveng et al 1994, Meurer et al 1995).  The IRAS data were taken
from the PSCz Catalogue (Saunders et al 2000), where careful attention was paid to IRAS fluxes
for extended galaxies.  For comparison 
there are 1538 galaxies satisfying the 60 $\mu$m, B and velocity constraints, so about 10 $\%$ of
these galaxies have 2000 $\AA$ data available.  The choice of 2000 $\AA$
allows an accurate estimate of the rate of formation of young massive stars, uncontaminated by light
from older stars.  Wide-beam measurements are needed because the galaxies are nearby and highly
extended. 
 
Figure 1 shows a plot of 
f(2000 $\AA$) versus f(60 $\mu$m), where f = $\nu S_{\nu}$.  The filled circles are for galaxies
with wide-beam uv data.  The crosses show the location of galaxies satisfying the $V_o$, B and S(60)
constraints but with uv data available only from IUE (Kinney et al 1993).
It is clear that the small beam of IUE seriously undersamples the uv emission from many of these
nearby galaxies.  In their study of local starburst galaxies, Meurer et al (1999) exclude
galaxies larger than 4' and argue that even though the IUE beam is 12 times smaller than
this, there is no significant undersampling.  Their sample includes galaxies beyond the velocity 
limit used here but 35 out of 57 galaxies in their sample (their Table 1) fall in the larger
sample illustrated in Fig 1.  Their physical justification for their claim that there is no
significant undersampling is that for the starburst galaxies they are considering, the uv
emission is concentrated to the nucleus of the galaxy. Since undersampling is clearly
a problem for many of the IUE galaxies in Fig 1, 
in this study I will exclude galaxies for which there is only IUE data in the uv.
 
Most of the galaxies in Fig 1 have values of $f(60 \mu m)/f(2000 \AA)$ in the range 0.3-10,
but this still represents a significant spread in the characteristic extinction in
nearby galaxies and it is clear that it is a crude approximation to apply a single average extinction correction 
to all galaxies, as has been done in most studies of the star-formation history to date.

\begin{figure}
\epsfig{file=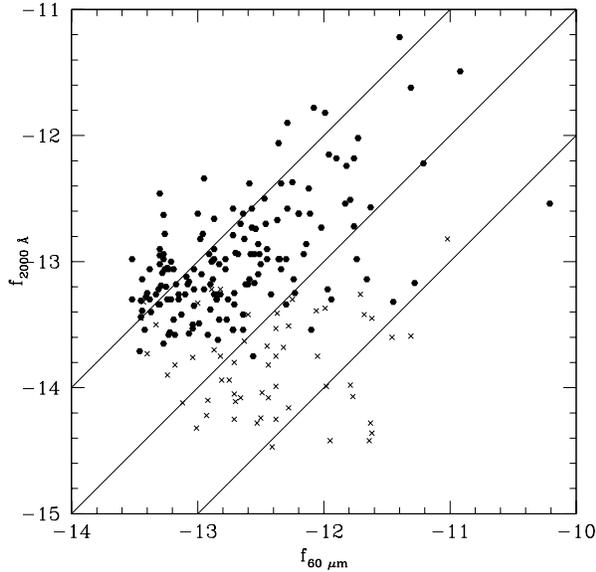,angle=0,width=8cm}
\caption{
f(2000 $\AA$) versus f(60 $\mu$m), in $W m^{-2}$, for bright nearby galaxies.  The broad-beam 
ultraviolet data (filled circles, 147 galaxies) are taken 
from Donas et al 1987, Buat et al 1987, Deharveng et al 1994, Meurer et al 1995.  The 60 $\mu$m
data are from IRAS PSCz catalogue (Saunders et al 2000).
The (narrow-beam) IUE data of Kinney et al (1993) are shown as crosses.  The IUE
data are displaced below the broad-beam data because of under-sampling of many of these
large nearby galaxies by IUE.
 The full list of galaxies and their properties is given at
http://astro.ic.ac.uk/$\sim$mrr/photz}
\end{figure}

Because we want to make a census of star-formation in all galaxies, not just those classified
as 'starbursts', my local sample is much broader than that of Meurer et al (1999).  Because of
the IRAS 60 $\mu m$ selection, it is still not completely representative of the local universe.
Elliptical and lenticular galaxies are undersampled, comprising only 8 $\%$ of the sample.
However there is probably almost no star-formation going on in the missed galaxies.  A useful
diagnostic diagram for the galaxies is a plot of the far infrared colour $lg_{10} S(100)/S(60)$
versus 60 $\mu m$ luminosity (Fig 2).  It can be seen that the majority of the galaxies have
$lg_{10} S(100)/S(60) > 0.3$ and $lg L_{60} < 10$, characteristic of 'cirrus' emission
(Rowan-Robinson and Crawford 1989).  The majority of the emission is from cool dust in
the interstellar medium, illuminated by the stellar radiation field of the whole galaxy.
Rather few galaxies have $lg_{10} S(100)/S(60) \sim 0$, characteristic of dust emission
dominated by a starburst.  Note that in calculating luminosities, velocities have been corrected for 
effects of peculiar motions using the flow model of Rowan-Robinson et al (2000).
The 35 galaxies from the Meurer et al (1999) sample which satisfy our $V_o$, B and S(60)
constraints are shown in Fig 2 as crosses.  As expected, they lie mainly in the region
$lg_{10} S(100)/S(60) < 0.3$.

The ideal way to check whether cirrus emission is dominant in the far infrared is to study
the submm fluxes.  Dunne et al (2000) have mapped a sample of 104 local galaxies at
850 $\mu m$, which does not, however, have any overlap with the present sample.
Figure 3 shows $lg (S(100)/S(60))$ versus $lg (S(60)/S(850))$ for the Dunne et al
sample.  The cirrus and starburst templates used by Rowan-Robinson (2001) have values of
$lg (S(60)/S(850))$ of 1.40 for the cirrus and 2.55 for the starburst template.  
Assuming a simple 2-component mixture, for galaxies with $lg (S(60)/S(850)) <$ 1.8 more 
than 50$\%$ of the far infrared
emission is from cirrus, and for $lg (S(60)/S(850)) <$ 2.0, more than 30$\%$.  Thus from
Fig 3 we can conclude that for galaxies with $lg (S(100)/S(60)) > $ 0.3, cirrus dominates,
and is still a strong contributor for galaxies with $lg (S(100)/S(60))$ = 0.2-0.3.
Note that the Dunne et al (2000) sample is biased against cooler galaxies (those with
$lg (S(100)/S(60)) >$ 0.4.

\begin{figure}
\epsfig{file=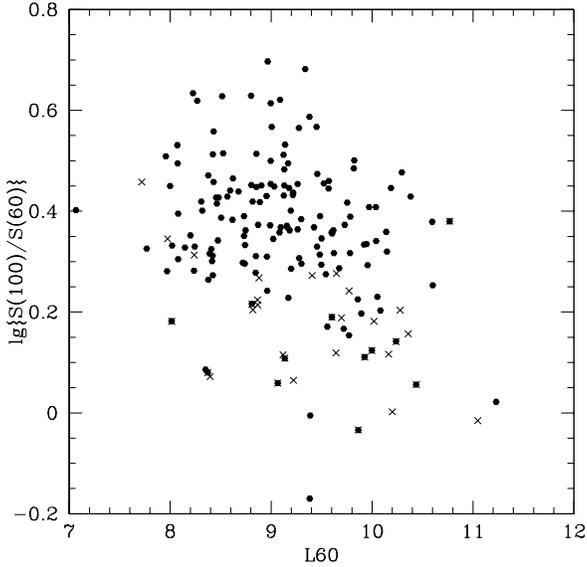,angle=0,width=8cm}
\caption{ The far infrared flux ratio  
$lg_{10} (S(100)/S(60))$
versus the 60 $\mu$m luminosity, $L_{60}$, for galaxies in the local sample of the present paper. The colours (most $>$ 0.3)
and luminosities (most $< 10^{10}$) are characteristic of galaxies
dominated by infrared cirrus.
Crosses denote galaxies from the Meurer et al (1999) study of starburst galaxies.}
\end{figure}

\begin{figure}
\epsfig{file=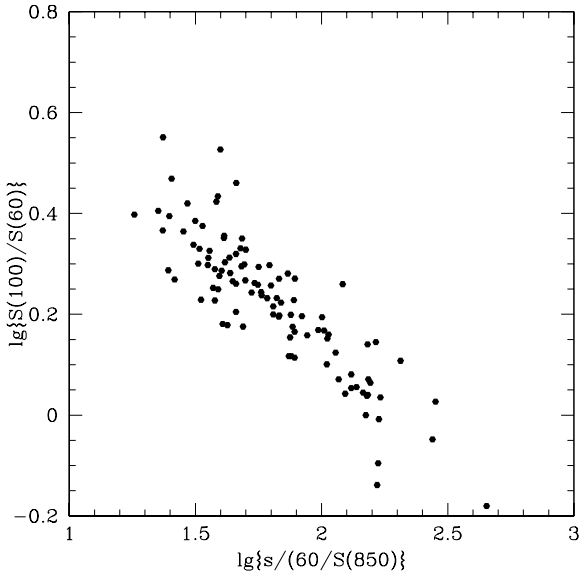,angle=0,width=8cm}
\caption{ The far infrared flux ratio  
$lg_{10} (S(100)/S(60))$
versus $lg_{10} (S(60)/S(850))$ for galaxies from the study of Dunne et al (2000).  Galaxies
with   $lg_{10} (S(60)/S(850)) <$ 1.8 are 
dominated by infrared cirrus.}
\end{figure}

I now use the far ir (60 $\mu$m)/far uv (2000 $\AA$) flux-ratio, and the [4400-2000] colour to estimate the characteristic
extinction in these galaxies.  

I define $L_{fir} = \int_{1 \mu m}^{1000 \mu m}  L_{\lambda} d\lambda = L_{60} bolc_{60}$,

where $bolc_{60}$ is the bolometric correction at 60 $\mu m$,

and  $IRX_{2000} = L_{fir}/L_{2000}$.

As in Meurer et al (1999) I assume that the far infrared emission is due to thermal emssion by dust 
heated by the radiation it absorbs so (cf their eqn (3))

\medskip

$IRX_{2000} =  [L_{Ly\alpha} + \int_{912 \AA}^{5 \mu m} L_{\lambda} (1-10^{-0.4 A_{\lambda}}) d\lambda]$

$ / (L_{2000,o} 10^{-0.4 A_{2000}})$.  (1)

\medskip

This corresponds to absorption by a screen of dust.  For optical depths not $>>$ 1 (as turns out to be the case for the 
majority of galaxies in this sample) this equation will also apply
approximately if dust and illuminating stars are mixed, with the derived optical depth representing an
average value.  Where the interpretation becomes complex, and requires a proper radiative transfer
treatment, is when in at least some directions the optical depth becomes high (Witt and Gordon 2000).  This is known to be
the case for starbursts, for example.  

At this point I diverge from the calculation of Meurer et al (1999) in two ways.  Firstly I evaluate (1)
exactly, rather than using the approximation given by Meurer et al's eqn (4).  I assume that 

\medskip

$L_{Ly\alpha}  =  (1-10^{-0.4 A_{1216}}) L_{Ly\alpha,o}$, 

\medskip

with $L_{Ly\alpha,o} = 0.2 L_{1216 \AA}$, corresponding to Case B recombination.  In practice many starburst
galaxies have $Ly\alpha$ emission very much weaker than the Case B prediction, due to either the aging of
the starburst (Valls-Gabaud 1993) or multiple scattering in the Ly$\alpha$ line
(Charlot and Fall 1993), but this would have only a small effect on the curves in Fig 5.

Secondly I assume initially
that we are dealing primarily with cirrus, absorption by interstellar dust in the galaxy.  This
affects the bolometric correction applied in the far infrared (Meurer et al 1999 use a correction
based on the Rigopoulou et al (1996) observations of ultraluminous IRAS galaxies).  It also affects
the wavelength range over which the integral in eqn (1) should be evaluated.  Meurer et al (1999)
assume that illumination is only by young massive stars, so that they only consider uv radiation
in the integral in eqn (1) (in their case the chosen uv wavelength is 1600 $\AA$).  For cirrus
the integration needs to be carried out from 912 $\AA$ into the near ir (I have set 5 $\mu m$ as
the long wavelength limit).  This results in uv bolometric corrections that depend strongly on
galaxy type (see Table 1).  The seds I have used in the uv-opt-near ir are those derived as photometric
redshift templates by Rowan-Robinson (2003), based on Yoshi and Takahara (1988) seds for normal galaxies
and Kinney et al (1993) data for a starburst sed.

\begin{figure*}
\epsfig{file=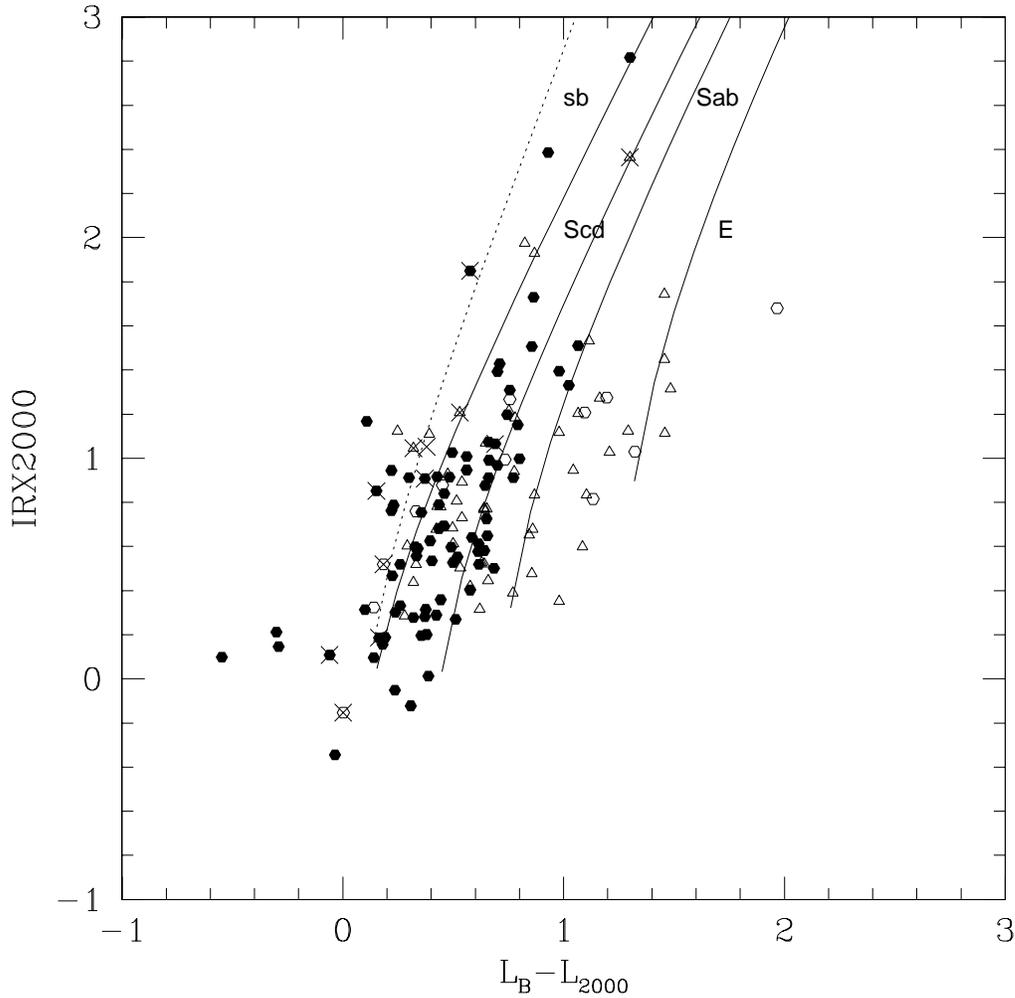,angle=0,width=14cm}
\caption{Infrared excess, IRX2000 = $L_{fir,bol}/L{2000}$ versus $L_B - L_{2000}$ for 141 local galaxies
with large-beam uv (2000 $\AA$) measurements.  Filled circles: de Vaucouleurs type T = 5-10, open triangles:
T = 1-4, open circles T $<$ 1.  Solid curves are calculated from eqn (2) using seds for E, Sab, Scd and
sb (starburst) galaxies from Rowan-Robinson (2002), assuming normal interstellar dust.  Broken curve
is calculated for sb galaxies using the Calzetti (1999) extinction law.}
\end{figure*}

\begin{table*}
\caption{Bolometric corrections used in present study.  Far infrared templates are taken
from Rowan-Robinson (2001).  Ultraviolet to near infrared templates are taken from
Rowan-Robinson (2002).}
\begin{tabular}{lllll}
\hline
 & $(1-1000 \mu m)/60 \mu m$ & $(912 \AA - 5 \mu m)/2000 \AA$ & $(912 \AA - 5 \mu m)/1600 \AA$ & $(912 \AA - 0.4 \mu m)/1600 \AA$ \\
\hline
cirrus fir compt & 3.25 & & & \\
& & & & \\
starburst fir compt & 1.92 & & & \\
& & & & \\
E & & 73.6 & & \\
& & & & \\
Sab & & 15.35 & & \\
& & & & \\
Scd & & 5.97 & & \\
& & & & \\
sb & & 4.00 & 3.90 & 1.81 \\
\hline
\end{tabular}
\end{table*}

Figure 4 shows a plot of $IRX_{2000}$ against $L_B - L_{2000}$ for the galaxies in my local sample,
with different symbols for the de Vaucouleurs et al (1976) type T (the 11 galaxies with no 
Hubble classification are omitted from Fig 4 for clarity though they are included in the subsequent 
statistical analyses).  The solid lines are calculated from eqn (1) for 4 sed types: starburst,
Scd, Sab and E, assuming normal Galactic dust.  
Although only 3 galaxies in the sample are classified as ellipticals (7 are lenticulars), the locus nicely defines
the right-hand edge of the distribution.
Also shown for the starburst case is the
corresponding curve for the Calzetti (1999) extinction law.  

Two features are apparent in Fig 4.  There is a broad horizontal spread of the galaxies, consistent
with the assumption that illumination by older stars becomes increasing important as we go to
earlier type galaxies.  This spread is broadly consistent with the de Vaucouleurs type.  However
there are clear cases of galaxies to which an early type is assigned but for which their location in this 
diagram shows they are more consistent with a starburst sed. The galaxies from the Meurer et al (1999) sample
are indicated by large crosses: they do indeed lie to the left of the distribution, consistent
with their being starbursts.

Secondly the distinction between the predictions of Galactic and Calzetti extinction laws are not great
in this diagram.  There is no need to assume for the quiescent star-formation typical of local
galaxies that anything other than a normal Galactic extinction law is involved.

To see whether the starburst sample of Meurer et al (1999) really does support a Calzetti, rather than a Galactic,
extinction law, I have reanalyzed their Fig 1, which is a plot of $IRX$ 
$= L_{fir,bol}/L_{optuv,bol}$ against the far ultraviolet
slope $\beta$, where $f_{\lambda} \propto \lambda^{\beta}$. Because the uv slope for Galactic dust
varies rather strongly with wavelength, one has to be precise about the definition of $\beta$
used for the theoretical curves.  The values for the galaxies were determined by fitting
to 10 bins sampled from IUE data in the wavelength range 1250-2500 $\AA$ (Meurer et al 1999).  
For the theoretical curves I have therefore defined $\beta = (lg_{10} (f_{2500}/f_{1250})/lg_{10} (2500/1250)$.
Figure 5 shows my version of their Fig 1, with my assumed starburst sed, and eqn (1) solved
exactly for the wavelength range 912 $\AA$ to 5 $\mu m$.  Again there is very little difference
in the predictions of the two extinction laws. This type of analysis on its own can not be
used to conclude that anything other than normal
Galactic dust is involved.  However if we look at IUE spectra in the Kinney et al (1993) atlas
for individual starburst galaxies with high values of IRX (and hence high inferred E(B-V) ), we 
find several cases where starburst galaxies do not show the strong 2200 $\AA$ feature that
would be implied by their E(B-V) if they were being extinguished by a screen of normal
Galactic dust.  Kinney et al (1993) draw attention to the case of NGC7552: NGC3690, NGC 4194
and NGC5135 are other examples of galaxies which have at most a weak 2200 $\AA$ feature,
despite having inferred E(B-V) values $>$ 0.4.
For starburst galaxies we therefore at the very least have to consider more elaborate dust 
geometries with, for example, uv radiation facing lines of sight with both moderate and
very high extinction.  Gordon et al (1997) and Witt and Gordon (2000) argue that from studies of a range of geometries
that the lack of the 2200 $\AA$ feature must be intrinsic to the dust extinction curve.
However Granato et al (2000) are able to reproduce the Calzetti 'extinction curve' and the
the observed distribution in Fig 5 from their starburst simulations, using Milky Way dust,
provided the age of the starbursts is less than 50 million years.
I return to the issue of starburst galaxies in section 3.

\begin{figure}
\epsfig{file=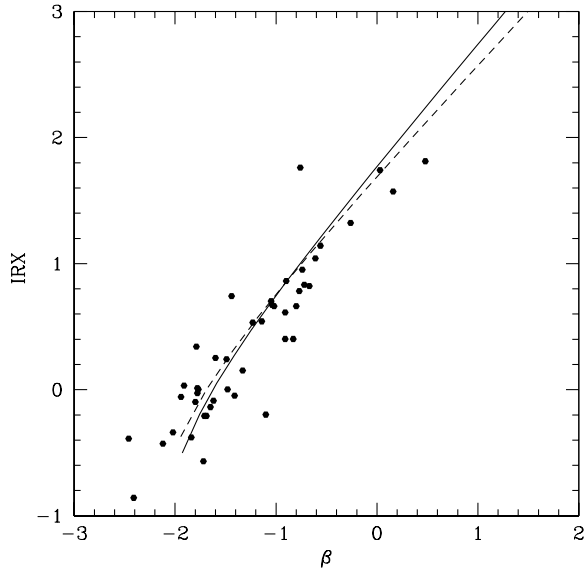,angle=0,width=8cm}
\caption{Infrared excess, IRX = $L_{fir,bol}/L{optuv,bol}$ versus uv slope $\beta (0.125-0.25 \mu m)$
for local starburst sample of Meurer et al (1999).  Solid curve is calculated from eqn (1) for normal
interstellar grains.  Broken curve is the corresponding locus for a Calzetti (1999) extinction law.}
\end{figure}

To derive an extinction for each galaxy in the local sample, which I characterize by E(B-V), I need to assume an 
sed type.  From Fig 4 it is clear that the de Vaucouleurs type offers only a broad indication of horizontal
location in the diagram and would lead to substantial errors in E(B-V) if used to specify an sed.  Instead
I make a grid of E(B-V) loci (Fig 6) and determine E(B-V) for each galaxy by interpolation.  The E(B-V) 
values are given in the full data table at $http://astro.ic.ac.uk/\sim mrr/photz$.  
The median value of E(B-V) is 0.14 and the mean value is 0.16.  These translate to values for $A_V, A_B, A_{2000}$ of 
0.43, 0.57, 1.20 for the median and 0.50, 0.67, 1.37 for the mean.  These values are broadly in agreement
with the range of extinctions found by Xu and Buat (1995) and Buat and Xu (1996).
Figure 7 shows the correlation between
E(B-V) and de Vaucouleurs type T.  All the galaxies with E(B-V) $>$ 0.4 also have
$lg (S(100)/S(60)) <$ 0.2, so are dominated by starbursts.   
For these galaxies, it is unlikely that
the derived value of E(B-V) applies to the overall interstellar medium of the galaxy. Since only a small
part of the 60 $\mu m$ flux will be due to cirrus emission, eqn (1) will overestimate the extinction of the
interstellar medium.  A further complication is that while much of the uv radiation in the starburst may be
completely absorbed by dust in the associated dense molecular cloud, some may leak out and suffer only 
partial extinction.  For such galaxies we really need a full radiative transfer treatment with non-spherical
geometry, but in the next section I attempt a first-order treatment.

The mean locus estimated by de Vaucouleurs et al (1976), from variations
of optical colours with inclination angles, are surprisingly good, particularly as similar analyses
were used by Disney et al (1989), Valentijn (1990), Phillips et al (1991), Valentijn (1994), to reach
the conclusion (clearly false from the present analysis, as also found by Xu and Buat 1995)
that most local spiral galaxies are completely optical thick in the visible across their whole disks.  

Figure 8 shows the correlation of E(B-V) with 60 $\mu m$ luminosity.  As suggested by several authors (Wang
and Heckman 1996, Buat et al 1999, Sullivan et al 2001) there is a trend of increasing E(B-V) with far infrared 
luminosity.  As we shall see in section 3, much of this effect is due to the increasing preponderance of a 
starburst component as $L_{60}$ increases, rather than to any genuine increase in the optical depth of
the interstellar medium in galaxies with luminosity.

Since the seds used in calculating the curves in Fig 4 from eqn (1) are empirical templates fitted to observed seds, 
they should strictly be dereddened by the average extinction in an iterative procedure, before a final value for
E(B-V) can be estimated.  The effect of this
dereddening moves the curves in Fig 4 to the left and downwards (except the curve for E galaxies, which is
unchanged).  The net effect on the estimated E(B-V) is in fact very small, so this correction has not been
applied here.

\begin{figure}
\epsfig{file=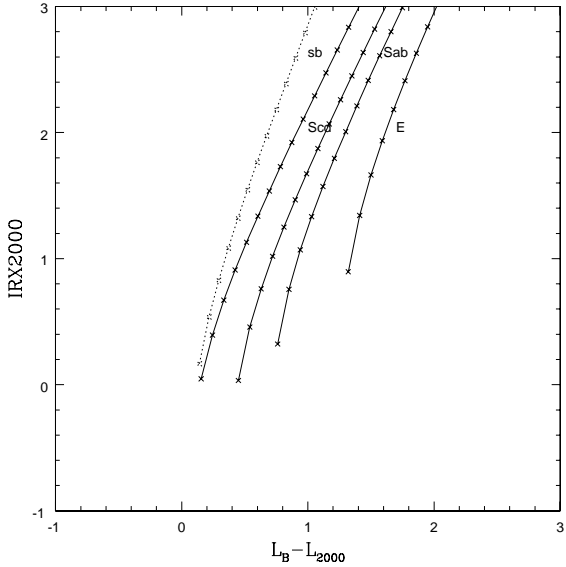,angle=0,width=8cm}
\caption{Same as Fig 4, with theoretical models marked with values of E(B-V), starting at bottom with 0.05,
and increasing in steps of 0.05.}
\end{figure}


\begin{figure}
\epsfig{file=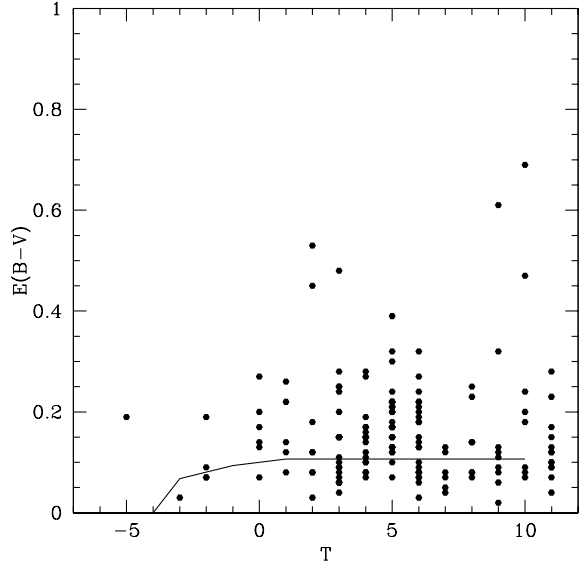,angle=0,width=8cm}
\caption{ E(B-V), calculated as in section 3, versus de Vaucouleurs type T.  The loci of mean values 
estimated by de Vaucouleurs et al 1976 are also shown.}
\end{figure}

\begin{figure}
\epsfig{file=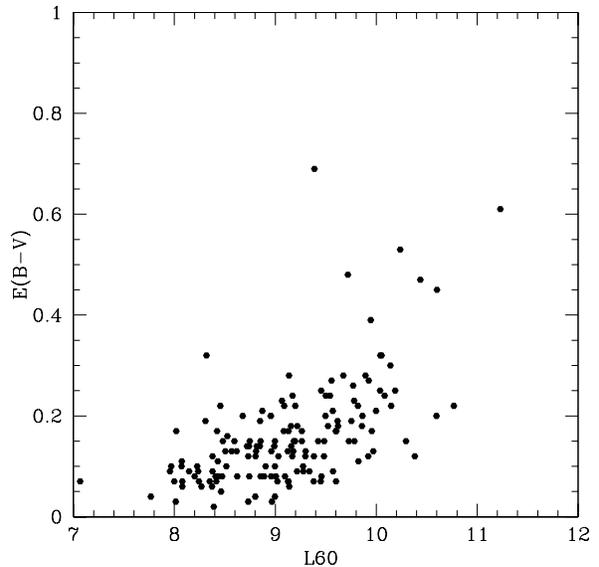,angle=0,width=8cm}
\caption{ E(B-V), calculated as in section 3, versus 60 $\mu m$ luminosity.  There is a trend of 
increasing extinction with luminosity.}
\end{figure}

\begin{figure}
\epsfig{file=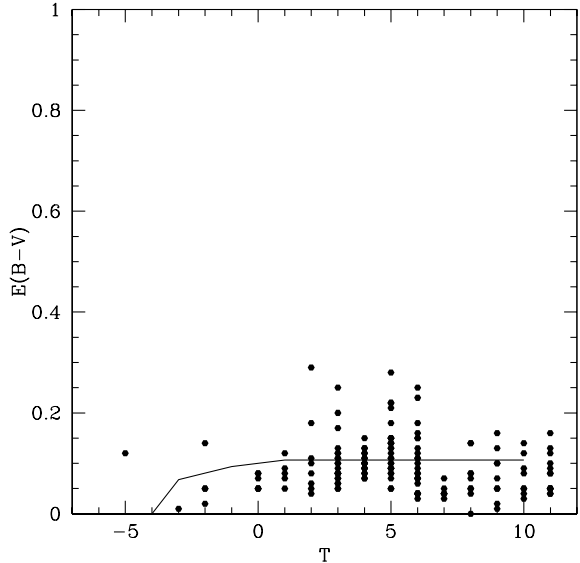,angle=0,width=8cm}
\caption{ E(B-V), estimated after subtraction of a starburst component in the far infrared (section 4), 
versus de Vaucouleurs type T.  The loci of mean values estimated by de Vaucouleurs et al 1976
are also shown.}
\end{figure}

\begin{figure}
\epsfig{file=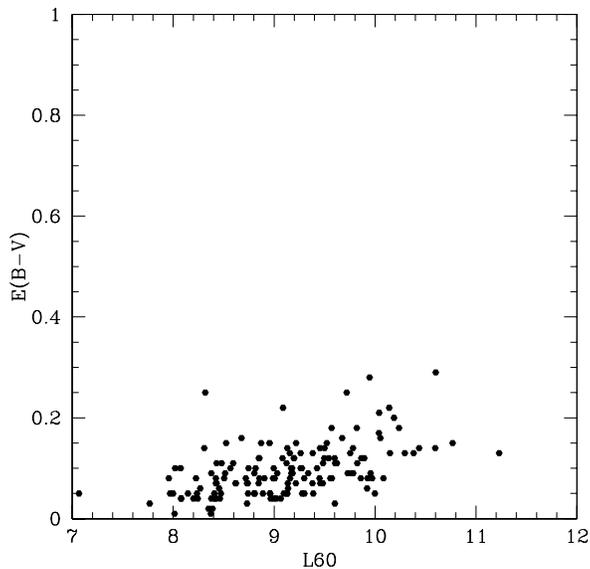,angle=0,width=8cm}
\caption{ E(B-V), estimated after subtraction of a starburst component in the far infrared (section 4), 
versus 60 $\mu m$ luminosity.  The trend of increasing extinction with luminosity is less steep.}
\end{figure}

\begin{figure}
\epsfig{file=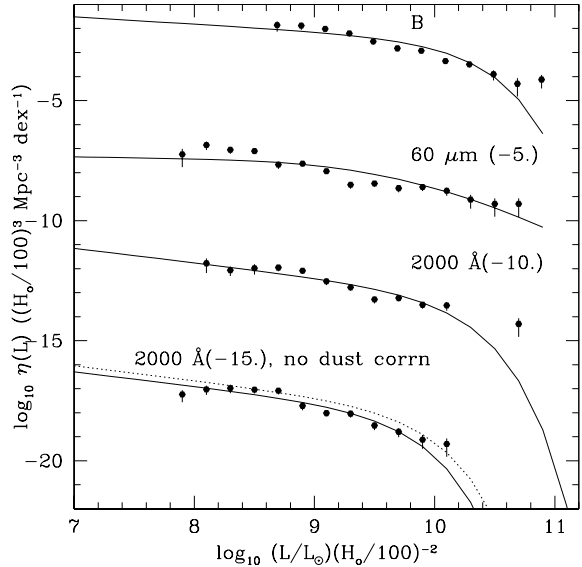,angle=0,width=8cm}
\caption{Luminosity function at B, 60 $\mu$m, 2000 $\AA$ corrected for extinction, and 2000 $\AA$ without
correction for extinction, for the
sample of 147 bright nearby galaxies, assuming effective area surveyed is 0.6 ster.  
The latter three plots have been displaced downwards by -5, -10 and -15 for clarity.  The B and 2000 $\AA$ data
have been fitted with Schecter functions and the 60 $\mu$m data have been fitted with the Saunders et al (1990)
function.}
\end{figure}

\section{Separation of cirrus and starburst far infrared components}

Rowan-Robinson and Crawford (1989) showed that the far infrared emission can be separated into optically thin
cirrus and optically thick starburst components, using IRAS colours.  To first order we can assume that for
the starburst component, none of the uv radiation escapes and this is the approach I adopt here.  I assume 
that the observed uv radiation is from stars spread through the galaxy and that the cirrus component is that 
proportion of the light from these stars which has been absorbed in the ism.  For the far infrared 'starburst' 
component I assume that the corresponding uv light has been completely absorbed by dust.

I use the ratio S(100)/S(60) to 
define $\alpha$, the fraction of the 60 $\mu m$ flux contributed by cirrus, using the starburst and cirrus templates
of Rowan-Robinson (2001) .  Specifically
\medskip

$\alpha$  =  (S(100)/S(60)-0.955)/2.651  (2)
\medskip

In the E(B-V) calculation I now replace S(60) by $\alpha$S(60).  The resulting E(B-V) values are generally lower and the 
tail at higher values disappears.  The median E(B-V) is now 0.09 and the mean is 0.010.
The revised E(B-V) for NGC7552 is 0.18, so the lack of a 2200 $\AA$ feature in the IUE data of Kinney et al (1993)
is, in this interpretation, unsurprising.

The agreement of these corrected E(B-V) values as a function of galaxy type T with the de Vaucouleurs mean values
(Fig 9) is now even better.   The slope of the correlation of E(B-V) with 60 $\mu m$ luminosity (Fig 10) is reduced compared 
to Fig 7, to
\medskip

E(B-V) $\sim$ 0.05 (lg$ L_{60} $- 7.0).       (3)
\medskip

In fact the evidence for a correlation of E(B-V) with $L_{60}$ is now not particularly strong, based on slightly higher 
E(B-V) values for galaxies with $lg_{10} L_{60} > 10$.

Although this separation into optically thin and completely optically thick dust components is almost certainly
an oversimplification, it does give a self-consistent picture and this is the approach I will use in the next section
to calculate the local uv luminosity-density and star-formation rate.

\section{2000 $\AA$ luminosity function and local uv luminosity-density}
For the 147 galaxies of my local sample I have derived the luminosity functions at 60 $\mu$m, B and 2000
$\AA$, the latter corrected for extinction by dust as described in section 3.  The 60 $\mu$m and B-band luminosity 
functions can be used to tell us whether the sample is
representative of all galaxies and allow the effective area surveyed to be estimated.
The upper two panels of figure  11 show fits to the luminosity functions at B and 60 $\mu$m, assuming the effective
area covered is 0.6 ster, and using the Schecter ($M_B = -19.5, \phi_* = 0.014, \alpha=1.3$, Loveday et al 1992) and Saunders et al
(1990) ($log L_* = 8.45, \sigma = 0.771, \alpha = 1.09, C_* = 0.028$) luminosity functions , respectively. 
The uncertainty in this estimate of the effective area covered is about 10 $\%$.
 Although the Saunders and Schecter luminosity 
functions give by no means a perfect fit to the 60 $\mu$m and B-band data, respectively, I conclude 
that this sample is representative of nearby bright
galaxies and hence that a fair estimate of the local 2000 $\AA$ luminosity density
can be derived. A further test is to compare the 2000 $\AA$ luminosity function, uncorrected
for the effects of extinction, with that derived for a slightly more distant sample ($z \sim 0.15$) 
by Treyer et al (1998).  This is shown in the bottom panel of Fig 11, with the solid curve showing the
best fit Schecter function with $\alpha$ = 1.6, and the broken curve showing the function derived
by Treyer et al (1998), assuming an area of 0.6 sr.  The agreement is excellent and the slight shift of
the latter curve to higher luminosities is entirely consistent with the rate of luminosity evolution
derived in numerous studies.  Thus apart from perhaps in the very highest luminosity bins, my sample does 
not appear to show any bias compared with true unbiassed surveys like that of Treyer et al (1998).
The faint-end slope of 1.6 at 2000 $\AA$ is supported not only by the work of Treyer et al (1998), but also by the
high redshift samples of Steidel et al (1999) and the uv luminosity function estimates of Rowan-Robinson (2003)
derived from photometric redshifts.

A Schecter function fit to the dust-corrected 2000 $\AA$ luminosity function is shown
in the third panel of Fig 11 (parameters are: $log L_* = 9.82, \phi_* = 0.0051, \alpha = 1.6$ ).  
The resulting luminosity density is  $10^{7.87} h L_{\odot} Mpc^{-3}$  and the 
corresponding star formation rate, using a conversion factor  $\dot{\phi_*} = 10^{-9.80} L_{2000}$
(Madau 1998, based on a Salpeter IMF and Bruzual and Charlot starburst model)
 is $10^{-1.63 \pm 0.06} h M_{\odot} yr^{-1} Mpc^{-3}$.  
To this must be added the contribution of the optically thick starburst components.  I estimate
this from the 60 $\mu m$ luminosity-density of Saunders et al (1989), using their estimate for the
'warm' component, assuming that essentially all the ultraviolet radiation is converted to far infrared
emission, and that the conversion factor is  $\dot{\phi_*} = 10^{-9.66} L_{60}$,
(Rowan-Robinson et al 1997, modified to take account of the revised starburst models by Madau (1998)).  This gives an
additional  $10^{-2.44} h M_{\odot} yr^{-1} Mpc^{-3}$, for a total of
 $10^{-1.57 \pm 0.06} h M_{\odot} yr^{-1} Mpc^{-3}$.  The optically thick starburst phase contributes
14 $\%$ of the total star-formation rate in local galaxies.  Correcting for the optically thick
starburst component, as in section 3, reduces the net estimate of the star-formation rate by a factor of 2.

This final corrected value agrees well with the estimate given by Gronwall (1998) from $H{\alpha}$,
$10^{-1.59 \pm 0.05} h M_{\odot} yr^{-1} Mpc^{-3}$ but is 
a factor 1.8 higher than the estimate given  
by Gallego et al (1995) based on $H{\alpha}$ data, $10^{-1.82 \pm 0.2}$.

Sullivan et al (2000) have given results from a balloon-borne uv survey of nearby galaxies,
which extends the results of Treyer et al (1998).  Their estimate of the star formation rate
at z = 0-0.3, corrected for dust extinction,
$10^{-1.21 \pm 0.05} h M_{\odot} yr^{-1} Mpc^{-3}$, is higher than that determined
locally, partly due to a higher extinction correction,
but is not inconsistent when account is taken of the strong variation of star-formation
rate with redshift.

There has been some discussion in the literature about whether H$\alpha$ or far-uv continuum flux
provides the most reliable estimate of the star-formation rate (eg Buat et al 2002, Bell 2003).  The key
issue is accounting for the effects of extinction.  Extinction estimates derived from Balmer line
ratios tend to be more than a factor of 2 higher than those estimated here.  The reason for this
is that much of the H$\alpha$ emission arises from a photosphere surrounding the highly
extinguished starburst region, so values of $A_V \sim$ 1 tend to be obtained. 
Extinction-corrected H$\alpha$ estimates are therefore likely to be less accurate than far infrared
estimates, for the highly extinguished component.  To disentangle the high and low optical
depth components, and the effect of older stellar populations, we need FUV, B-band, 60 and 100
$\mu m$ fluxes.  

The correction from FUV and 60 $\mu m$ (and H$\alpha$) luminosity-densities to star-formation rate has a strong
dependence on the assumed IMF and star-formation scenario.  Buat et al (2002) has given some
illustrations of this dependence for FUV and H$\alpha$ estimates (their table 4) 


\section{Conclusions}

(1)  I have used a sample of local ($V_o < 5000 km/s$) galaxies with uv and far ir data to estimate
the extinction in the interstellar medium of the galaxies.  In most of the galaxies  
( those with $log_{10} (S(100)/S(60)) > 0.2$) the far infrared
emission is dominated by cirrus emission.   
A significant fraction of the illumination of the dust is due to older, lower mass stars, and
this becomes progressively more important for earlier type galaxies.  Excluding starburst
galaxies there is no evidence that anything other than normal Milky Way dust is involved.
The mean extinction is found to correspond to E(B-V) = 0.16, implying $A_V = 0.50, A_B = 0.67, A_{2000} = 1.37$.

(2)  For starburst galaxies, analysis of infrared excess versus uv colour diagrams using a screen model
does not support a Calzetti (1999) rather than Milky Way dust extinction law.  However for a few such galaxies detailed
IUE spectra do not show the 2200 $\AA$ feature that would be expected for Milky Way dust with the 
inferred values of E(B-V) from a screen model.  At the very least a more detailed radiative transfer
model is required for starburst galaxies, with non-spherically symmetric geometry and with high optical 
depths in some lines of sight.  Here the 60/100 $\mu m$ flux-ratio is used to separate the contributions
of (optically thin) cirrus and (optically thick) starburst components to the far infrared emission.  This
has the effect of reducing the mean E(B-V) for the cirrus component to 0.10.

(3)  The 60 $\mu m$ and B-band luminosity functions for the local sample show that
the sample is representative of the local universe and corresponds to 0.6 sr of sky.
I have used the 2000 $\AA$ luminosity function, corrected for the effects of extinction
to derive the local ultraviolet luminosity-density,
$10^{7.87 \pm 0.06} h L_{\odot} yr^{-1} Mpc^{-3}$,
and the global star-formation rate, $10^{-1.57 \pm 0.06} h M_{\odot} yr^{-1} Mpc^{-3}$,
including the contribution of the optically thick starburst phase, which has to be estimated from
the far infrared.
This value is consistent with  the values derived by Gronwall (1998)
from $H{\alpha}$ data, though higher than the estimate of Gallego et al (1996).
It is also consistent with the value derived by Sullivan et al (2000) from a uv survey of
galaxies with $<z>$ = 0.15, when allowance is made for the strong dependence of
star-formation rate with redshift.


\end{document}